\documentstyle{article}

\newcommand{\alphafive}{\alpha^{\scriptscriptstyle (5)}}

\newcommand{\alphathree}{\alpha^{\scriptscriptstyle (3)}}
\newcommand{\alphatwo}{\alpha^{\scriptscriptstyle (2)}}
\newcommand{\alphazero}{\alpha^{\scriptscriptstyle (0)}}
\newcommand{\alphanf}{\alpha^{\left(n_f\right)}}

\newcommand{\Psub}{{\scriptscriptstyle  P}}
\newcommand{\alphaP}{\alphanf_\Psub}
\newcommand{\GeV}{{\rm GeV}}
\newcommand{\msbar}{{\scriptscriptstyle {\rm \overline{MS}}}}
\newcommand{\ainv}{{a^{-1}}}
\newcommand{\lat}{{\rm lat}}
\newcommand{\be}{\begin{equation}}
\newcommand{\ee}{\end{equation}}
\newcommand{\bearray}{\begin{eqnarray}}
\newcommand{\eearray}{\end{eqnarray}}
\newcommand{\eq}[1]{Eq.~(\ref{#1})}
\newcommand{\order}{{\cal O}}
\newcommand{\PS}{\chi_b - \Upsilon}
\newcommand{\PSc}{\chi_c - \psi/\eta_c}
\newcommand{\SS}{\Upsilon^\prime - \Upsilon}
\newcommand{\qchi}{q_{\chi_b}}
\newcommand{\qups}{q_{\Upsilon}}
\newcommand{\qupsp}{q_{\Upsilon^\prime}}

\newcommand{\MSbar}{\overline{\rm MS}}
\newcommand{\eff}{{\rm eff}}
 
\title{Further Precise Determinations of $\alpha_s$ from Lattice QCD}

\author{
C.~T.~H.~Davies,$^a$ K.~Hornbostel,$^b$
G.~P.~Lepage,$^c$\\ P.~McCallum,$^a$
J.~Shigemitsu,$^d$ J.~Sloan $^e$\\[.4cm]
\small $^a$University of Glasgow and the UKQCD Collab., Glasgow, UK G12 8QQ \\
\small $^b$Southern Methodist University, Dallas, TX 75275 \\
\small $^c$Newman Laboratory of Nuclear Studies,
            Cornell University, Ithaca, NY 14853 \\
\small $^d$The Ohio State University, Columbus, OH 43210 \\
\small $^e$Dept. of Physics and Astronomy, Univ. of Kentucky,
Lexington, KY 40506-0055\\
\\ }

\date{March 1997}

\begin{document}
\maketitle

\begin{abstract}
We present a new determination of the strong coupling constant from
lattice QCD simulations. We use four different short-distance
quantities to obtain the coupling, three different (infrared) meson
splittings to tune the simulation parameters, and a wide range of
lattice spacings, quark masses, and lattice volumes
to test for systematic errors. Our final result consists of ten
different determinations of~$\alphathree_P(8.2\,\mbox{GeV})$, which
agree well with each other and with our previous results. The
most accurate of these, when evolved perturbatively to the $Z^0$~mass, gives
$\alphafive_\msbar(M_Z)\!=\!.1174\,(24)$. We compare our results with
those obtained from other recent lattice simulations.
\\ \\ Keywords: Strong Coupling Constant, Lattice QCD, Quarkonium, 
         Perturbation Theory
\end{abstract}

\pagebreak

\section{Introduction}
Precise measurements of the strong coupling constant~$\alpha_s$ are
important not only for strong-interaction phenomenology, but also in
the search for new physics. Any discrepancy between low-energy and
high-energy determinations of this coupling could signal the
existence of supersymmetry or other phenomena beyond
the Standard Model.  No significant discrepancies have yet been 
observed~\,\cite{susy}; more stringent tests of the Standard Model
require further improvements in precision. 
In an earlier paper\,\cite{davies95} we showed that lattice 
simulations of quantum chromodynamics (QCD), when combined with the
very accurate experimental data on the $\Upsilon$~meson spectrum,
provide among the most accurate and reliable determinations
of~$\alpha_s$. 
$\Upsilon$'s probe the strong interactions at the relatively low
energies of 500--1000~MeV, where supersymmetry or other new
physics has little effect. Thus it is important to compare the
couplings obtained from lattice QCD with those obtained from high-energy
accelerator experiments, where effects due to a more fundamental
underlying theory would be much more important. 
And it is essential that these couplings be measured as accurately as
possible, with realistic estimates of the uncertainties involved. In
this paper we review our earlier determination of the coupling, and
update it to take
advantage of new results from third-order perturbation theory, as well as 
new simulations which substantially reduce some of our Monte Carlo
errors.  We also report on several
new simulations that further bound our systematic errors,
particularly with respect to contributions from quark vacuum
polarization.

As discussed in \cite{davies95}, there are two steps in our
determination of the coupling constant.  The first is to create a
numerical simulation that accurately mimics QCD dynamics. We do this
by tuning the bare masses and coupling in a lattice QCD simulation 
until it reproduces experimental results for the orbital
and radial excitations of $\Upsilon$ mesons.  We
use the $\Upsilon$~family because it is one of the few systems
for which both accurate simulations and accurate
experimental data are available.

Having tuned our simulation, the second step in our determination of
the coupling is to use the simulation to generate nonperturbative Monte
Carlo ``data'' for a variety of short-distance quantities. Comparison
with the perturbative expansions for the same quantities then fixes the
value of the QCD coupling constant. We
use the expectation values of small Wilson loops as our short-distance
quantities. These are very easy to compute in simulations. They are also
completely euclidean and very ultraviolet, and therefore largely free of
hadronization or other nonperturbative corrections. Finally, small
Wilson loops have very convergent perturbative expansions that are
known through second order for arbitrary~$n_f$, the number of
light-quark flavors, and through third order for~$n_f\!=\!0$.

In this paper we examine each of these steps in detail. We begin in
Section~2 by describing how we tune the simulation parameters. 
The most important of these for our analysis is the bare coupling constant,
or equivalently the lattice spacing, used in the lattice QCD
Lagrangian. The number and masses of light quarks entering through
vacuum polarization is also important; we present new simulation
results that bear on these parameters. In Section~3 we describe
several different determinations of $\alpha_\msbar$ using different
Wilson loops.  Each of these sections deals extensively with potential
systematic errors.  Finally, in Section~4 we summarize our results and
discuss future directions.

\section{Tuning of the Simulation}
\subsection{Procedure}
Given a lattice spacing~$a$, the QCD parameters that determine
$\Upsilon$~properties are the bare coupling constant~$g_\lat$ in the
lattice Lagrangian, the bare mass~$M^0$ of the constituent
$b$~quarks, and the bare masses~$m_q^0$ of the light quarks that
enter through quark vacuum polarization.  Only the 
$u$, $d$ and $s$ quarks are light enough to contribute to vacuum
polarization appreciably.
These parameters all vary with the lattice spacing. In a simulation, they
must be tuned so that physical quantities computed in the simulation
agree with the corresponding experimental values. The tuning
procedure is much simpler, and therefore more reliable, if one
uses physical quantities that are very sensitive to one of the
parameters and insensitive to the others.

Our main interest in this paper is the coupling constant, and so we
are particularly careful in tuning the bare coupling.
We use the mass splittings between radial and
orbital excitations of the $\Upsilon$ for this purpose. These splittings
are ideal since they are almost completely insensitive to the
$b$-quark mass.  The spin-averaged mass
splittings between 1P and 1S levels, and 2S and 1S levels are observed
experimentally to vary by only a few percent between the $\Upsilon$
and $\psi$ systems, even though $b$ quarks are roughly three times heavier
than $c$ quarks. This striking insensitivity to the mass of the
constituents is an accident, but is confirmed by simulations for 
a range of masses near the $b$~mass.

These splittings are also quite insensitive to the masses of the light
quarks.  These contribute through vacuum polarization, and
affect hadronic masses in two ways.  First, they allow decays
to multi-hadronic final states; mixing with these states shifts
the masses of the original hadrons. 
$\Upsilon$ decay rates are typically .1\% or less of the mass
splittings, and the states we examine are all far below the
$B\overline{B}$~threshold. Thus we may ignore such effects in our 
analysis.  The second effect of vacuum
polarization is to renormalize the gluonic interactions between the
constituents of the hadron. The typical momentum $\qups$ exchanged 
between the $b$ quarks in an~$\Upsilon$ is from
$.5$ to 1~GeV.  This is small compared to the $c$,
$b$~and $t$~quark masses, and we may ignore their contribution to
vacuum polarization.  In contrast, the $u$, $d$~and $s$~quarks are 
effectively almost massless at these energies and
must be included in a realistic simulation.  At the same time, because 
their masses are small relative to~$\qups$, our simulation results
depend only weakly on their exact values.

For sufficiently small masses, the dependence
of an $\Upsilon$ mass splitting should be linear\,\cite{san-diego}:
\be \label{m-dep}
\Delta M \approx \Delta M^0\,\left\{ 1 + {\rm constant}\times
\sum_{u,d,s}\frac{m_q^0}{\qups} + \cdots \right\},
\ee
where the renormalized 
$s$~mass is 50--100~MeV\,\cite{s-mass-papers}, and the $u$~and
$d$~masses are 20~or 30~times smaller and therefore negligible. 
It is very costly to simulate lattice QCD with
realistic $u$~and $d$~masses. Here that is unnecessary. 
The simple dependence of $\Delta M$ on~$m_q^0$ means that we 
obtain realistic results if we set all three light-quark masses equal to
$m_{\rm eff}^0\equiv m_s^0/3$, which generates the same
correction to~$\Delta M$ as two massless quarks and a strange
quark. Thus
$m_\eff\!=\!15$--30\,MeV, and \eq{m-dep} suggests that the
dependence on light-quark masses is a few percent or less of the total
mass splitting, comparable to the Monte Carlo statistical
errors in our analysis.\footnote{It is conceivable that the linear
term in \eq{m-dep}, which is due to chiral symmetry breaking, is
strongly suppressed for tiny mesons such as the $\Upsilon$, and
becomes nonleading. Then the dependence on $m_\eff$ would be
quadratic, with the correct value for $m_\eff\!=\!m^0_s/\sqrt{3}$.  
The sensitivity to $m_\eff$ would then be far smaller and probably 
negligible for our analysis.}

There are several other properties of the $\Upsilon$~system that make
it ideal for tuning the bare coupling. These mesons are essentially
nonrelativistic; the use of a nonrelativistic effective
action\,\cite{nrqcd-papers} to
exploit this allows a large portion of the spectrum to be computed
efficiently and
precisely\,\cite{davies-upsilon,davies-psi}. They are
physically small\,---\,three or four times smaller than light-quark
hadrons\,---\,and so do not suffer from finite-volume errors even on
modestly sized lattices. Finally, we have detailed phenomenological
quark models that are well-founded theoretically and that give us
unprecedented control over systematic errors.

In addition to the bare coupling constant, we must also tune the bare
masses of the $b$ quark and of the light quarks. We tune the bare
$b$-quark mass~$M^0$ by requiring that the $\Upsilon$ mass in the
simulation has its correct value of 9.46~GeV.  Ref.~\cite{davies94}
presents a detailed discussion.  The light-quark masses are tuned
until the pion and kaon masses are correct. As discussed, we
need only the $s$-quark mass, as we set all $n_f\!=\!3$ light-quark
masses to $m_s^0/3$.

Finally we note that it is customary in tuning lattice simulations to
switch the roles of the lattice spacing and the bare coupling constant.
Rather than choose a lattice spacing and then tune the bare
coupling constant to its correct value, it is far simpler to choose a
value for the bare coupling constant~$g_\lat$, and then {\em compute}
the corresponding lattice spacing $a$ using simulation results. All
explicit dependence on the spacing can be removed from the
simulation code by expressing dimensionful quantities in units of
$a$ or~$a^{-1}$.  The spacing is then not needed as an input to
the code, but is specified implicitly through the input value for
$g_\lat$, or equivalently through $\beta\!\equiv\!6/g_\lat^2$. 
We determine $a$ from the $\Upsilon$~mass splittings~$\Delta M$.  
The simulation produces these in
the dimensionless combination $a\,\Delta M$; to obtain $a$, we divide
by the experimentally measured values for $\Delta M$.

The lattice spacing is a crucial ingredient in our determination of
the renormalized coupling
$\alpha_s$.  As we discuss in Section~\ref{alpha-section}, the
short-distance quantities we study specify~$\alpha_s(q^*)$ for a
specific value of $a q^*$. The expectation value of the
$a\!\times\!a$~Wilson loop, for example, gives
$\alpha_s(q^*)$ for $q^*\!=\!3.4/a$.  For this to be useful,
we need to know $q^*$, and therefore $a^{-1}$, in physical units such
as GeV.  Consequently, the next section focuses on
how precisely we are able to determine the lattice
spacing corresponding to a given value of~$\beta$.

\subsection{Results: $a^{-1}$ Determination }
Our lattice simulations used the standard Wilson action for the
gluons, and the staggered-quark action and the Hybrid\ Molecular\
Dynamics algorithm for the light quarks. We employed a nonrelativistic
formulation of quark dynamics (NRQCD) for the
$b$ quarks\,\cite{nrqcd-papers,davies-upsilon,davies-psi}. The
$n_f\!=\!0$ gauge-field configurations used in our Monte Carlo
calculations were provided by G.~Kilcup and his collaborators
($\beta\!=\!6,6.4$)\,\cite{kilcup}, J.~Kogut
($\beta\!=\!6$)\,\cite{kogut}, and by
the UKQCD collaboration ($\beta\!=\!5.7,6.2$)\,\cite{ukqcd}. The
$n_f\!=\!2$ configurations are from the SCRI Lattice Gauge Theory
Group and their colleagues in the HEMCGC collaboration
($\beta\!=\!5.6$)\,\cite{hemcgc}, and from the
MILC collaboration ($\beta\!=\!5.415,5.47$)\,\cite{milc}. 
Unfortunately, we were unable to obtain
configurations with $n_f\!=\!3$ light-quark flavors, which is the
correct number for $\Upsilon$~physics. Consequently, we performed
complete analyses for $n_f\!=\!0$ and $n_f\!=\!2$ and 
extrapolated our results to $n_f\!=\!3$.  The extrapolation was
the last step of our analysis, and is described in
Section~\ref{alpha-section}. 

As discussed above, we use mass splittings in the $\Upsilon$~system to
determine the lattice spacing. Specifically,
we use two different mass splittings to make two
independent determinations of the lattice spacing. One is the
splitting \mbox{$\Delta M(\SS)$} between the $\Upsilon^\prime$ and the $\Upsilon$,
and the other is the splitting \mbox{$\Delta M(\PS)$} between the spin
average of the $\chi_b$ mesons and the $\Upsilon$.
These can be measured accurately in a
simulation\,\cite{davies-upsilon,davies-psi}, and are known
very accurately from experiments.  Table~\ref{ainv-results}
summarizes the parameters used in
our main simulations and the results for these two splittings.
Our most reliable results are based on the $\beta\!=\!6$ and~5.6
simulations.  We use results from the other simulations,
including those for the splitting between the spin-averaged $\chi_c$~mesons 
and the spin average of the $J/\psi$ and $\eta_c$~mesons,
to calibrate systematic errors.
Our $\beta\!=\!6.2$ result agrees with that in~\cite{cambridge}.

\begin{table}
\begin{center}
\begin{tabular}{lcclclcl}
\hline \\
$\beta$ & $n_f$ & $a\,m_\eff^0$ & $a\,M^0_q$ & splitting & $a\,\Delta M$
&
$a\,\Delta M_g$ & $a^{-1}$ (GeV) \\[1.5ex]
\hline 
6.0  & 0 & -- &1.71 & $\PS$ & .174\,(3) & $-.004$ & 2.59\,(4)(4)(0)\\
&&&1.80&&.174\,(12) \\
&&&2.00&&.173\,(10) \\ \\

6.0  & 0 & -- &1.71 & $\SS$ & .242\,(5) & $-.001$ & 2.34\,(5)(3)(0) \\
&&&1.80 &  & .239\,(11) \\
&&&2.00 & & .235\,(11) \\ \\

5.6 & 2 & .010 & 1.80 & $\PS$ & .185\,(5) & $-.005$ & 2.44\,(7)(4)(7)\\
    &   & .025 &  &  & .200\,(12) &   \\ \\

5.6 & 2 & .010 &1.80 & $\SS$ & .239\,(10) & $-.002$ & 2.38\,(10)(3)(10)\\
    &   & .025 & &  & .262\,(13) &  \\ \\

5.7 & 0 & -- & 0.80 & $\PSc$ & .383\,(10) & $-.009$ & 1.22\,(3)(18)(0)\\
&&& 3.15 & $\PS$ & .326\,(6) & $-.015$ & 1.41\,(4)(4)(0)\\ \\

5.415& 2 & .0125& 0.80 & $\PSc$ & .359\,(14) & $-.008$ &
       1.30\,(5)(20)(5)\\ 
&&& 2.80 & $\PS$ & .323\,(10) & $-.017$  &
      1.44\,(6)(4)(6)\\ \\

5.47 & 2 & .05 & 0.80 & $\PSc$ & .335\,(15) \\ 
&&& 2.8 & $\PS$ & .307\,(12) \\ \\ 


6.2 & 0 & -- & 1.22 & $\PS$ & .124\,(5) & $-.001$ & 3.58\,(15)(5)(0) \\
    &   &    &      & $\SS$ & .175\,(8) & $-.0003$ & 3.22\,(15)(5)(0) \\ \\

6.4 & 0 & -- & 1.00 & $\PS$ &  .107\,(16) & $-.002$  &
4.19\,(63)(6)(0) \\
\hline
\end{tabular}
\end{center}
\caption{Simulation results for meson mass splittings~$a\,\Delta M$ 
and inverse lattice spacings~$a^{-1}$, in GeV, for a range of
couplings~$\beta$, light-quark masses~$m_q^0$ and heavy-quark
masses~$M_q^0$. The gluonic $a^2$~corrections $a\,\Delta M_g$ 
shown are added to $a\,\Delta M$
to obtain the corrected splitting. The error estimates 
for~$a^{-1}$ are for statistical errors, $a^2$~and $v^4$~errors, and
errors in the light-quark mass, respectively. 
Experimental values for $\Delta M$
are .563~GeV for $\SS$, .440~GeV for $\PS$, and .458~GeV for $\PSc$.}
\label{ainv-results}
\end{table}

Several factors contribute to the uncertainty in our determination of
the lattice spacing. We used the lattice NRQCD formalism to simulate
heavy-quark dynamics\,\cite{nrqcd-papers}, and included all relativistic
corrections through~$\order(v^2)$ and all finite lattice-spacing
corrections through~$\order(a^2)$. The leading finite-$a$ error is due
to~$\order(a^2)$ errors in the gluon dynamics. We estimate this effect
using perturbation theory\,\cite{davies95}, which indicates that only
$S$~states are affected and that our measured $S$-state energies
should be shifted by
\be
a\,\Delta M_g = \frac{3}{40}\,\left(a\,M_q\right)^2\,a\,\Delta M_{\rm
HFS},
\ee
where $\Delta M_{\rm HFS}$ is the hyperfine spin splitting of the state and
$M_q$~is the heavy-quark mass.  We assume 1.5~GeV and 5~GeV for $c$
and $b$~quarks, respectively.  The corrections we used are
listed in Table~\ref{ainv-results}, 
as are our final values for the inverse lattice
spacing~$a^{-1}$.
We allow for a systematic error of~$\pm\Delta M_g/2$ in~$\Delta M$
when computing the error in~$a^{-1}$, although our analysis in~\cite{davies95}
suggests a much smaller uncertainty. 
Note that $\Delta M(\SS)$ is almost unaffected by this correction.
Relativistic corrections of
order~$v^4$ are most likely negligible for the~$\Upsilon$ 
since the $v^2$~corrections,
which we include, shift our mass splittings by less than~10\%; we
include a systematic uncertainty of~$\pm1$\% for this.
The $v^4$~corrections are certainly larger for~$\psi$'s, where, for
example, the $J/\psi$--$\eta_c$ splitting is a $v^2$~effect and 25\%
of the $\PSc$~splitting. This suggests $v^4$~errors could be of
order~$\pm6$\% for $\psi$'s. Recent simulations\,\cite{trottier}
indicate that
certain spin-dependent $v^4$ terms can shift levels by as much as
60\,MeV, which is 15\% of the splitting. We include a
systematic uncertainty 
of~$\pm15$\% for $v^4$~errors in the $\psi$ splitting.

Our simulations confirm that the $b$-quark mass has very little effect
on either of the $\Upsilon$~splittings. The $\beta\!=\!6$ results show
that a 17\% change in~$M^0$ leads to changes of only a few percent in
the $\SS$ and $\PS$ splittings. (Note that the statistical errors in the
splittings for different $M^0$'s are correlated.  Consequently, the
statistical errors in the differences between the splittings are
somewhat smaller than those for any individual splitting.) Since
we determine $M^0$ to within 6\%\,\cite{davies94}, the resulting
uncertainty in the determination of the lattice spacing is probably no
more than a percent, which is much smaller than the statistical errors.

Uncertainties in the light-quark mass can also affect our lattice
spacing determination. In our $\beta\!=\!5.6$ simulations we expect
$a\,m_s^0$ to be somewhere in the range~0.01--0.02. This can be
inferred from the dependence of the pion mass on $m_q^0$, 
and allows for uncertainties due to quenching and finite-$a$
errors. Thus we want light-quark masses $a\,m_q^{0}\!=\!a\,m_{\rm
eff}^0$ in the range~.003--.006. We have simulation results for
$a\,m_\eff^0=.01$ and~.025. By fitting formula~(\ref{m-dep}) to 
these results we find that $a\,m_q^0=.01$ should give the correct
result to within~4\%, which equals  our statistical error. 
The correct range of light-quark masses in our $\beta\!=\!5.415,5.47$
simulations is roughly $a\,m_\eff^0=.005$--.015. We have simulation
results for $a\,m_\eff^0=.0125$ and~.050, and again the 6-7\% shift
caused by changing $m_\eff$ is roughly the same as our statistical
errors for both $\psi$~and 
$\Upsilon$~splittings.\footnote{To 
compare mass splittings at $\beta\!=\!5.415$ with
those at $\beta\!=\!5.47$ one needs the expectation value of the plaquette
at each beta; see the following section. From the plaquette values one finds that
the lattice spacing at the larger beta is about 12\% smaller. Since
the $a\,\Delta M$'s are only 5--6\% smaller at the larger beta, the
splittings~$\Delta M$ themselves are actually about 6--7\% larger for the
larger mass.}  Note that $\psi$'s should be more
sensitive to small quark masses than $\Upsilon$'s since they are
roughly twice as large; we saw no evidence for this in our simulations.
These results all indicate that the $m_\eff$~dependence is too small 
compared to our statistical errors to allow an accurate
measurement.\footnote{This insensitivity to $m_\eff$ is because
$m_\eff$ is so small in our simulations. Our $n_f\!=\!0$ simulations are
equivalent to $m_\eff\!=\!\infty$ and give results that are quite
different from $n_f\!=\!2$. So shifts would become apparent, even with
our statistics, for sufficiently large $m_\eff$.} This also means that
the tuning errors
associated with $a\,m_\eff$ are no larger than our statistical errors,
and so we take our statistical errors as a measure of the uncertainty
due to this parameter.

We checked for finite-volume errors by computing the charmonium splittings
using lattices that are 1.5\,fm and 3.0\,fm per side.  We observed no 
difference, indicating that these errors are smaller than the 2\%
statistical errors in these tests. The lattices we used at $\beta\,=\,6$
and~5.6 are both $16 a\!\approx\!1.35$\,fm per side; the $\Upsilon$'s
are half the size of the $\psi$'s, with a radius of about~.2\,fm.  We
therefore expect finite volume errors in our mass splittings that are 
substantially less than~2\%. 

We estimated the electromagnetic shifts of the $\Upsilon$ masses 
using a potential model.  For individual mesons, we found mass shifts of
approximately 1~MeV, with smaller shifts for the splittings between them.
These are too small to affect our result.

Our final values for $a^{-1}$'s are listed in Table~\ref{ainv-results}, obtained
by dividing the experimental values for the splittings $\Delta M$ by
the corrected Monte Carlo simulation results $a\,\Delta M + a\,\Delta
M_g$. The error estimates for the $a^{-1}$'s 
include statistical errors in $a\,\Delta M$, as well as systematic errors
associated with the finite-$a$ correction $a\,\Delta M_g$,
$v^4$~corrections, and the light-quark mass~$m_q^0$. Other systematic
errors are negligible. 

Perhaps the most striking feature of these simulation results is the
significant disagreement at $\beta\!=\!6$ between $\ainv$ computed
using the $\SS$~splitting and that computed using the $\PS$~splitting.
Taking proper account of correlations, this disagreement is 
five standard deviations: our simulation gives 1.43(3) for the ratio
of these splittings, rather than the experimental value of 1.28.  
Thus the $\beta\!=\!6$ simulation is inconsistent with experiment.
This is because in this simulation, in contrast to nature, 
$n_f\!=\!0$; there is no light-quark vacuum polarization.
The disagreement is much smaller when $n_f\!=\!2$, as is apparent in the 
$\beta\!=\!5.6$ data.  And, as we will demonstrate, it disappears
completely when we extrapolate~$n_f$ to three. 

As expected, using an incorrect value for $n_f$ leads to
inconsistencies such as the one found in our $\beta\!=\!6$ simulation.
Perturbation theory, though not justified at the
momenta relevant for these systems, provides a qualitative explanation 
for this discrepancy.  The centrifugal barrier
makes the average separation between the quarks in the $P$ state $\chi_b$
larger than for the $S$ state
$\Upsilon$ or $\Upsilon^\prime$, as is familiar from hydrogen or
positronium.  As a result, the typical exchanged momentum for $\chi_b$
quarks, $\qchi$, is smaller than $\qupsp$.  The perturbative binding energy
is given by
$\alpha_s^2(q) C_F^2 M_b/16$, with $q=\qupsp$ for 
$\Upsilon^\prime$ and $\qchi$ for $\chi_b$. Since $\qchi\!<\!\qupsp$, the
$\chi_b$ is more tightly bound. However, for $n_f\!=\!0$, this effect is
exaggerated, as
$\alphazero_s(q)$ increases more quickly than $\alphathree_s(q)$ with
decreasing $q$. Thus, for $n_f\!<\!3$, \mbox{$\Delta M(\PS)$} should be
underestimated relative to \mbox{$\Delta M(\SS)$}, as is observed.  Fitting to
data would then require a larger $\ainv$ for \mbox{$\Delta M(\PS)$} than for
\mbox{$\Delta M(\SS)$}.

\begin{figure}
\begin{center}
\setlength{\unitlength}{.02in}
\begin{picture}(95,100)(40,-50)

\put(50,-50){\line(0,1){85}}
\multiput(48,-40)(0,20){4}{\line(1,0){4}}
\multiput(49,-40)(0,10){7}{\line(1,0){2}}
\put(47,-40){\makebox(0,0)[r]{$-40$}}
\put(47,-20){\makebox(0,0)[r]{$-20$}}
\put(47,0){\makebox(0,0)[r]{$0$}}
\put(47,20){\makebox(0,0)[r]{$20$}}
\put(47,35){\makebox(0,0)[r]{MeV}}


\put(63,-5){\makebox(0,0)[l]{$h_{\rm b}$}}

\put(70,-0.8){\circle*{3}}
\put(75,-2.9){\circle{3}}
\put(75,-2.9){\line(0,1){1.2}}
\put(75,-2.9){\line(0,-1){1.2}}
\multiput(90,-40)(3,0){7}{\line(1,0){2}}
\put(110,-40){\makebox(0,0)[l]{$\chi_{\rm b0}$}}
\put(97,-24){\circle*{3}}
\put(97,-23){\line(0,1){1}}
\put(97,-25){\line(0,-1){1}}
\put(102,-34){\circle{3}}
\put(102,-34){\line(0,1){5}}
\put(102,-34){\line(0,-1){5}}

\multiput(90,-8)(3,0){7}{\line(1,0){2}}
\put(110,-8){\makebox(0,0)[l]{$\chi_{\rm b1}$}}
\put(97,-8.6){\circle*{3}}
\put(102,-7.9){\circle{3}}
\put(102,-7.9){\line(0,1){2.4}}
\put(102,-7.9){\line(0,-1){2.4}}

\multiput(90,13)(3,0){7}{\line(1,0){2}}
\put(110,13){\makebox(0,0)[l]{$\chi_{\rm b2}$}}
\put(97,10.1){\circle*{3}}

\put(102,11.5){\circle{3}}
\put(102,11.5){\line(0,1){2.4}}
\put(102,11.5){\line(0,-1){2.4}}
\end{picture}
\end{center}
 \caption{NRQCD simulation results for the spin structure of the
  lowest-lying $P$ states.
  Dashed lines indicate experimental values for the triplet
$P$ states.   
Masses are relative
to the spin-averaged state. Results are from a simulation with $n_f\!=\!0$
(filled circles) and from one with $n_f\!=\!2$ (open circles),
using $\ainv\!=\!2.4~\GeV$ for both.  
The errors shown are statistical; systematic errors are within about
5~MeV.}
\label{spect-splits}
\end{figure}
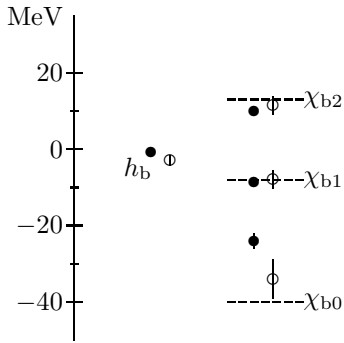

We end this section by displaying in Figures~\ref{spect-fig}
and~\ref{spect-splits} results from the $\beta\!=\!6$ and~5.6
simulations for  several of the low-lying excitations and spin
splittings, compared with experimental values.  
The agreement is excellent and supports the reliability of our
simulations. We emphasize that these are calculations from first
principles; our approximations can be systematically improved.  The
only inputs are the Lagrangians describing gluons and quarks, and the
only parameters are the bare coupling constant and quark masses. In
particular, these simulations are {\em not\/} based on a
phenomenological quark potential model.

\begin{figure} 
\begin{center}
\setlength{\unitlength}{.02in}
\begin{picture}(130,150)(5,920)
\thicklines
\put(10,935){\line(0,1){125}}
\multiput(8,950)(0,50){3}{\line(1,0){4}}
\multiput(9,950)(0,10){10}{\line(1,0){2}}
\put(7,950){\makebox(0,0)[r]{9.5}}
\put(7,1000){\makebox(0,0)[r]{10.0}}
\put(7,1050){\makebox(0,0)[r]{10.5}}
\put(7,1065){\makebox(0,0)[r]{GeV}}



\put(27,920){\makebox(0,0)[t]{${^1\rm{}S}_0$}}
\put(25,943.1){\circle*{3}}
\put(30,942){\circle{3}}

\put(52,920){\makebox(0,0)[t]{${^3\rm{}S}_1$}}
\multiput(43,946)(3,0){7}{\line(1,0){2}}
\put(50,946){\circle*{3}}
\put(55,946){\circle{3}}

\multiput(43,1002)(3,0){7}{\line(1,0){2}}
\put(50,1004.1){\circle*{3}}
\put(50,1005.1){\line(0,1){0.2}}
\put(50,1003.1){\line(0,-1){0.2}}
\put(55,1003){\circle{3}}
\put(55,1004){\line(0,1){1.4}}
\put(55,1002){\line(0,-1){1.4}}

\multiput(43,1036)(3,0){7}{\line(1,0){2}}
\put(50,1060){\circle*{3}}
\put(50,1061){\line(0,1){11}}
\put(50,1059){\line(0,-1){11}}
\put(55,1039.1){\circle{3}}
\put(55,1039.1){\line(0,1){7.2}}
\put(55,1039.1){\line(0,-1){7.2}}

\put(92,920){\makebox(0,0)[t]{${^1\rm{}P}_1$}}

\multiput(83,990)(3,0){7}{\line(1,0){2}}
\put(90,987.6){\circle*{3}}
\put(95,989){\circle{3}}
\put(95,990){\line(0,1){0.2}}
\put(95,988){\line(0,-1){0.2}}

\multiput(83,1026)(3,0){7}{\line(1,0){2}}
\put(90,1038.7){\circle*{3}}
\put(90,1039.7){\line(0,1){1.4}}
\put(90,1037.7){\line(0,-1){1.4}}
\put(95,1023){\circle{3}}
\put(95,1023){\line(0,1){7.2}}
\put(95,1023){\line(0,-1){7.2}}

\put(120,920){\makebox(0,0)[t]{${^1\rm{}D}_2$}}
\put(120,1019.2){\circle*{3}}
\put(120,1020.2){\line(0,1){6}}
\put(120,1018.2){\line(0,-1){6}}
\end{picture}
\end{center}

 \caption{NRQCD simulation results for the spectrum of the
$\Upsilon$ system, including radial excitations.
  Dashed lines indicate experimental values for the triplet
$S$ states and for the
 spin average of the triplet $P$ states. The energy zero from
 simulation results is adjusted to give the correct mass to the
 $\Upsilon(1{^3S}_1)$. Results are from a simulation with $n_f\!=\!0$
(filled circles) and from one with $n_f\!=\!2$ (open circles),
using $\ainv\!=\!2.4~\GeV$ for both.  The errors shown are statistical;
systematic errors are of order 20\,MeV or less.} 
\label{spect-fig}
\end{figure}
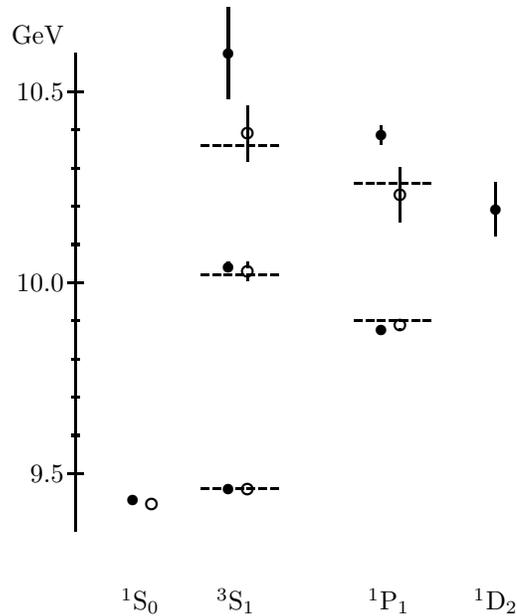

\section{Determination of the Renormalized Coupling}\label{alpha-section}

\subsection{The Coupling Constant from Wilson Loops}\label{wloop-section}

Having tuned the simulation, we performed Monte Carlo simulations
to generate ``data'' for a variety of short-distance quantities. 
We determined the coupling 
by matching the perturbative expansions for these quantities
to the nonperturbative Monte Carlo results.  For short-distance
quantities we chose the expectation values~$W_{m,n}$ of Wilson loop
operators.  In the continuum, 
\be
W_{m,n} \equiv \mbox{$\frac{1}{3}$}\,\langle\, {\rm ReTr}\,{\rm P}\,{\rm
e}^{-ig\oint_{n,m}
\!A\cdot dx}\, \rangle,
\ee
where ${\rm P}$~denotes path ordering, $A_\mu$~is the QCD vector
potential, and the integral is over a closed
${ma\!\times\!na}$~rectangular path. Loop operators for small paths are
among the most ultraviolet, and therefore most perturbative, objects that
can be studied in lattice QCD simulations. Unlike most other quantities
used to determine the QCD coupling, the loop operators are truly
short-distance quantities in euclidean space. There are no corrections
for hadronization, and nonperturbative effects are expected to be very
small. For example, the leading nonperturbative
contribution to $W_{m,n}$ due to condensates is probably from the gluon
condensate, with
\be
\delta W_{m,n}  = -\,\frac{\pi\,a^4\,(mn)^2}{36}\,
\langle\alpha_s\,F^2\rangle\, .
\ee
Most studies find that $\langle\alpha_s\,F^2\rangle$ is of
order .042~GeV$^4$~\cite{yndurain}.
Since $a^{-1}$~ranges from 1.2~to 4.2~GeV in our simulations, 
we expect condensate contributions to $-\ln W_{1,1}$, for example, to be
in the range of~.2--.01\%, much too small to be important here. 
When $n_f\!\ne\!0$ there are also contributions from
quark condensates, but these are suppressed by $\alpha_s^2$ and
so are probably even smaller. The tiny size of such effects make the
$W_{m,n}$ for small~$m$ and~$n$ ideal quantities for
determining the coupling in lattice QCD, particularly given the ease with
which they can be computed in simulations.

To obtain four independent determinations of the coupling, 
we used expectation values for the four
smallest loops on the lattice: the plaquette $W_{1,1}$,
$W_{1,2}$, $W_{1,3}$, and $W_{2,2}$. Each of these loop operators is
very different from the others; as different, for example, as various 
moments of a structure function.  Each is affected differently by
nonperturbative effects and higher-order uncalculated perturbative
corrections.  The contribution of the gluon condensate, for example, is
16~times larger for $W_{2,2}$ than for $W_{1,1}$. By comparing
results obtained from different loop operators we can bound such systematic
errors. 

Each of our expectation values has a perturbative
expansion of the form
\be
-\ln  W_{m,n}^{(n_f)}  = \sum_{i=1} c_i^{(n_f)}(m,n)\,
          (\,\alphanf_\Psub(q_{m,n})\,)^i \, ,
\ee
where $\alpha_\Psub$ is a new nonperturbative 
definition for the coupling constant 
introduced in our earlier paper\,\cite{davies95} to facilitate lattice
calculations.  The scale $q_{m,n}$ is the average gluon momentum in the
first-order contribution to $W_{m,n}$, computed directly from the
Feynman diagrams as described in\,\cite{lm,blm}. 

In Table~\ref{wloops-pth} we list the perturbative
coefficients through third order for $n_f\!=\!0$, and
through second order for $n_f\!=\!2$\,\cite{wloop-pth-ref}.
Unfortunately, the $n_f$~dependence of the third-order coefficients has not
yet been computed. Given that the second-order coefficients depend only
weakly on~$n_f$ by design~\cite{lm,blm}, it is likely that the
$n_f\!=\!0$ third-order coefficients are also good approximations
when~$n_f\!=\!2$. We assume this in our analysis, but when
estimating errors at $n_f\!=\!2$ we take the size of the entire
$n_f\!=\!0$ third-order contribution as an estimate of the uncertainty
due to $n_f$~dependence. When $n_f\!=\!0$, we estimate the truncation
error in perturbation theory to be of order $\alpha_\Psub^3(q_{m,n})$
times the leading order contribution.

\begin{table}
\begin{center}
\begin{tabular}{crrrrc}
\hline
loop & \multicolumn{1}{c}{$c_1$} & \multicolumn{2}{c}{$c_2$} &
\multicolumn{1}{c}{$c_3$} & $a\,q_{m,n}$\\
&& {$n_f\!=\!0$} & 
{$n_f\!=\!2$} & 
{$n_f\!=\!0$} \\ \hline
$-\ln W_{1,1}$ & 4.19 & $-4.98$ & $-5.57$ & 0 & 3.40 \\
$-\ln W_{1,2}$ & 7.22 & $-7.57$ & $-8.51$ & 2.6 & 3.07 \\
$-\ln W_{1,3}$ & 10.07& $-9.60$ & $-10.89$ & 5.3 & 3.01 \\
$-\ln W_{2,2}$ & 11.47& $-10.58$ & $-11.84$ & 11.1 & 2.65 \\ \hline
\end{tabular}
\end{center}
\caption{Coefficients for the perturbative expansions, in powers of
$\alpha_\Psub(q_{m,n})$, of small Wilson loops. Scale~$q_{m,n}$ is the
average momentum carried by the gluon in the first-order correction.} 
\label{wloops-pth}
\end{table}

Note that the plaquette $W_{1,1}$ has no third-order contribution. 
This is because the coupling~$\alpha_\Psub$ is defined in terms of the
plaquette\,\cite{davies95}; the absence of third and higher-order
corrections is merely a consequence of our conventions. Truncation errors
in the plaquette's expansion reappear when our coupling is converted to
more standard couplings, such as $\alpha_\msbar$:
 \bearray \label{msb-v}
 \alphanf_\msbar(Q) & = & \alphanf_\Psub(e^{5/6}\,Q) 
\,\left\{ 1 + 2\,\alphanf_\Psub/\pi \right. \\ \nonumber
& + & \left.X_\msbar\,(\,\alphanf_\Psub\,)^2
 + \order(\,(\,\alphanf_\Psub\,)^3\,) \right\}\, .
 \eearray
Here the third-order coefficient~$X_\msbar\!\approx\!0.95$ for
$n_f\!=\!0$\,\cite{luscher95}. The third-order coefficient is new since
our first paper. Unfortunately, the $n_f$~dependence of this
coefficient is not known. However, the variation of this coefficient
as $n_f$~goes to two or three is unlikely to be large. The
factor~$e^{5/6}$ in the scale is chosen to eliminate 
$n_f$~dependence in the second-order coefficient of the
expansion\,\cite{blm}, and therefore also removes much of the
$n_f$~dependence in third order. As above, we use the $n_f\!=\!0$ value
for $X_\msbar$ throughout our analysis, but when $n_f\!=\!2$ we take the
size of the entire third-order term as our estimate of the uncertainty
due to $n_f$~dependence.

The coupling $\alpha_\Psub$ was defined to coincide through second order
with the continuum coupling~$\alpha_V$ defined in \cite{blm,lm} from the
static-quark potential. The third-order correction to the static-quark
potential has recently been computed\,\cite{markus}, leading to
\be
\alphanf_V(Q) = \alphanf_\Psub(Q)\,\left\{ 1 +
X_V\,(\,\alphanf_\Psub\,)^2 + \cdots \right\} \, ,
\ee
where $X_V\!=\! 1.86 - .14\, n_f + X_\msbar$, which is $2.81$ for $n_f\!=\!0$. 
Note that this expansion has
infrared divergences in fourth-order and beyond, due to residual retardation
effects in the static quark potential\,\cite{appelquist}.

\subsection{Results: $\alpha_\Psub$ Determinations}
 
Monte Carlo simulation results for the expectation values of the Wilson
loop operators are summarized in Table~\ref{wloops-data}\,\cite{wloops-mc}. 
We also tabulate the values of $\alpha_\Psub(q_{m,n})$ obtained by matching
perturbation theory to Monte Carlo 
simulation results. The uncertainties quoted are
our estimates of the potential truncation errors in perturbation theory;
see  Section~\ref{wloop-section}.  The only other potential sources of
error are nonperturbative effects, and as discussed, these are almost
certainly negligible compared to truncation
errors. Finite-volume errors are much less than 1\% for such small loops.

\begin{table}
\begin{center}
\begin{tabular}{cccccc}
\hline \\
$\beta$ & $n_f$ & $a\,m_\eff^0$ & loop & M.C. value &
$\alphanf_\Psub(q_{m,n})$   \\[1.5ex]
\hline 
6.0 & 0 & -- & $-\ln W_{1,1}$ &.5214\,(0) &.1519\,(0) \\  
&&& $-\ln W_{1,2}$ & .9582\,(1)  &.1571\,(6)  \\
&&& $-\ln W_{1,3}$ & 1.3757\,(2) & .1584\,(6) \\
&&& $-\ln W_{2,2}$ & 1.6605\,(3) & .1657\,(8)  \\ 
\\
5.6 & 2 & .025& $-\ln W_{1,1}$ & .5719\,(0) & .1792\,(0)\\
&&  .010 & $-\ln W_{1,1}$ & .5709\,(0)& .1788\,(0)    \\
&&.010& $-\ln W_{1,2}$ & 1.0522\,(1) & .1828\,(30)   \\
&&.010& $-\ln W_{1,3}$ & 1.5123\,(2) & .1832\,(40)   \\
&&.010& $-\ln W_{2,2}$ & 1.8337\,(3) & .1907\,(80)  \\ 
\\
5.7 & 0 & -- & $-\ln W_{1,1}$  & .5995\,(0) & .1829\,(0)  \\ 
\\
5.415 & 2 & .0125& $-\ln W_{1,1}$ & .6294\,(0) & .2075\,(0) \\
 \\
 5.47 & 2 & .050& $-\ln W_{1,1}$ & .6134\,(0) & .1993\,(0)   \\ 
\\
6.2 & 0 & -- & $-\ln W_{1,1}$ & .4884\,(0) & .1398\,(0)  \\
\\
6.4 & 0 & -- & $-\ln W_{1,1}$ & .4610\,(0) & .1302\,(0)  \\ 
\hline
\end{tabular}
\end{center}
\caption{Expectation values of Wilson loop operators for small loops,
and the corresponding $\alpha_\Psub$'s for a variety of lattice QCD
parameters. The uncertainties listed for the expectation values are Monte
Carlo statistical errors.  Those listed for the
$\alpha_\Psub$'s are estimates of the truncation errors in perturbation
theory.}
\label{wloops-data}
\end{table}

The values for the various coupling constants in this table are all
different. This is because the coupling-constant scales $q_{m,n}$ are
different for each operator and for each parameter set. To compare
these results we must first evolve the running coupling constants to a
common scale. In Table~\ref{alphaP-evol-table} we present the
couplings evolved to 8.2\,GeV, which is the scale we chose
in~\cite{davies95}. To generate these values, we converted the
corresponding~$q_{m,n}$'s from units of~$\ainv$ to GeV using the lattice
spacings inferred from each of the $\Upsilon$ or $\psi$ mass splittings for
which we have 
simulation results. We then evolved the couplings to 8.2\,GeV by
numerically integrating the evolution equation for $\alpha_\Psub$. We used
the universal second-order beta function together with the $n_f\!=\!0$
third-order term for $\alpha_\Psub$. The $n_f$~dependence of the third-order 
beta function is unknown, but the entire third-order term generally has
negligible effect. This is especially true for our most important
results at $\beta\!=\!6$ and~5.6, since 8.2\,GeV was
chosen to be very close to the $q_{m,n}$'s and very little
evolution is required.

\begin{table}
\begin{center}
\begin{tabular}{cccclll}
\hline \\
$\beta$ & $n_f$ & $a\,m_\eff^0$ & loop &
\multicolumn{3}{c}{$\alphanf_\Psub(8.2\,{\rm GeV})$} 
\\ \cline{5-7}
&&&& \multicolumn{1}{c}{$\PS$}   & 
\multicolumn{1}{c}{$\SS$}  & 
\multicolumn{1}{c}{$\PSc$} \\[1.5ex]
\hline 
6.0 & 0 & -- & $-\ln W_{1,1}$    & .1552\,(10)(0) &  .1505\,(11)(0) \\
&&& $-\ln W_{1,2}$ & .1556\,(10)(6) & .1509\,(11)(6) \\
&&& $-\ln W_{1,3}$  & .1560\,(11)(6) & .1512\,(11)(6) \\
&&& $-\ln W_{2,2}$ & .1565\,(11)(8) & .1517\,(11)(8) \\
\\
5.7 & 0 & -- & $-\ln W_{1,1}$ & .1528\,(18)(0) & & .1465\,(61)(0) \\
 \\
6.2 & 0 & -- & $-\ln W_{1,1}$ & .1569\,(23)(0) & .1519\,(23)(0) \\
\\
6.4 & 0 & -- & $-\ln W_{1,1}$ &  .1515\,(67)(0) \\ 
\hline 
5.6 & 2 & .010 & $-\ln W_{1,1}$ & .1794\,(24)(0) & .1781\,(33)(0) \\
&&.010& $-\ln W_{1,2}$ & .1777\,(24)(30) & .1764\,(32)(30) \\ 
&&.010& $-\ln W_{1,3}$ & .1770\,(24)(40) & .1757\,(32)(40) \\
&&.010& $-\ln W_{2,2}$ & .1767\,(23)(71) & .1754\,(32)(71) \\
\\
5.415 & 2 & .0125& $-\ln W_{1,1}$ & .1748\,(34)(0) & & .1696\,(78)(0) \\
\hline
\end{tabular}
\end{center}
\caption{Values of $\alpha_\Psub(8.2\,{\rm GeV})$ from several
operators $W_{m,n}$ and a variety of tunings for QCD
simulations, with different $\beta$'s, $n_f$'s, and 
meson mass splittings used to fix $\ainv$. The two uncertainties listed are
due to uncertainties in the inverse lattice spacing, and to
truncation errors in the extraction of
$\alpha_\Psub$ using perturbation theory. }
\label{alphaP-evol-table}
\end{table}

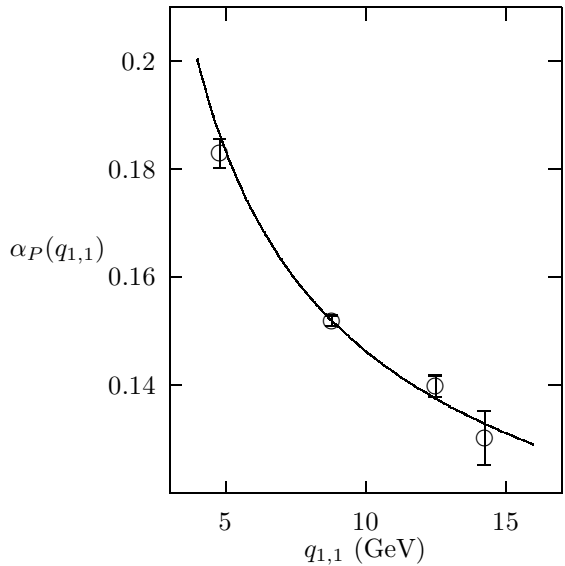
\begin{figure}
\begin{center}
\setlength{\unitlength}{0.240900pt}
\ifx\plotpoint\undefined\newsavebox{\plotpoint}\fi
\sbox{\plotpoint}{\rule[-0.200pt]{0.400pt}{0.400pt}}%
\begin{picture}(900,900)(0,0)
\font\gnuplot=cmr10 at 10pt
\gnuplot
\sbox{\plotpoint}{\rule[-0.200pt]{0.400pt}{0.400pt}}%
\put(220.0,283.0){\rule[-0.200pt]{4.818pt}{0.400pt}}
\put(198,283){\makebox(0,0)[r]{$0.14$}}
\put(816.0,283.0){\rule[-0.200pt]{4.818pt}{0.400pt}}
\put(220.0,453.0){\rule[-0.200pt]{4.818pt}{0.400pt}}
\put(198,453){\makebox(0,0)[r]{$0.16$}}
\put(816.0,453.0){\rule[-0.200pt]{4.818pt}{0.400pt}}
\put(220.0,622.0){\rule[-0.200pt]{4.818pt}{0.400pt}}
\put(198,622){\makebox(0,0)[r]{$0.18$}}
\put(816.0,622.0){\rule[-0.200pt]{4.818pt}{0.400pt}}
\put(220.0,792.0){\rule[-0.200pt]{4.818pt}{0.400pt}}
\put(198,792){\makebox(0,0)[r]{$0.2$}}
\put(816.0,792.0){\rule[-0.200pt]{4.818pt}{0.400pt}}
\put(308.0,113.0){\rule[-0.200pt]{0.400pt}{4.818pt}}
\put(308,68){\makebox(0,0){$5$}}
\put(308.0,857.0){\rule[-0.200pt]{0.400pt}{4.818pt}}
\put(528.0,113.0){\rule[-0.200pt]{0.400pt}{4.818pt}}
\put(528,68){\makebox(0,0){$10$}}
\put(528.0,857.0){\rule[-0.200pt]{0.400pt}{4.818pt}}
\put(748.0,113.0){\rule[-0.200pt]{0.400pt}{4.818pt}}
\put(748,68){\makebox(0,0){$15$}}
\put(748.0,857.0){\rule[-0.200pt]{0.400pt}{4.818pt}}
\put(220.0,113.0){\rule[-0.200pt]{148.394pt}{0.400pt}}
\put(836.0,113.0){\rule[-0.200pt]{0.400pt}{184.048pt}}
\put(220.0,877.0){\rule[-0.200pt]{148.394pt}{0.400pt}}
\put(45,495){\makebox(0,0){$\alpha_{P}(q_{1,1})$}}
\put(528,23){\makebox(0,0){$q_{1,1}$ (GeV)}}
\put(220.0,113.0){\rule[-0.200pt]{0.400pt}{184.048pt}}
\put(264,795){\usebox{\plotpoint}}
\multiput(264.58,788.71)(0.497,-1.786){49}{\rule{0.120pt}{1.515pt}}
\multiput(263.17,791.85)(26.000,-88.855){2}{\rule{0.400pt}{0.758pt}}
\multiput(290.58,698.10)(0.497,-1.362){51}{\rule{0.120pt}{1.181pt}}
\multiput(289.17,700.55)(27.000,-70.548){2}{\rule{0.400pt}{0.591pt}}
\multiput(317.58,625.82)(0.497,-1.142){49}{\rule{0.120pt}{1.008pt}}
\multiput(316.17,627.91)(26.000,-56.908){2}{\rule{0.400pt}{0.504pt}}
\multiput(343.58,567.57)(0.497,-0.911){51}{\rule{0.120pt}{0.826pt}}
\multiput(342.17,569.29)(27.000,-47.286){2}{\rule{0.400pt}{0.413pt}}
\multiput(370.58,518.90)(0.497,-0.810){49}{\rule{0.120pt}{0.746pt}}
\multiput(369.17,520.45)(26.000,-40.451){2}{\rule{0.400pt}{0.373pt}}
\multiput(396.58,477.29)(0.497,-0.693){49}{\rule{0.120pt}{0.654pt}}
\multiput(395.17,478.64)(26.000,-34.643){2}{\rule{0.400pt}{0.327pt}}
\multiput(422.58,441.62)(0.497,-0.592){51}{\rule{0.120pt}{0.574pt}}
\multiput(421.17,442.81)(27.000,-30.808){2}{\rule{0.400pt}{0.287pt}}
\multiput(449.58,409.80)(0.497,-0.537){49}{\rule{0.120pt}{0.531pt}}
\multiput(448.17,410.90)(26.000,-26.898){2}{\rule{0.400pt}{0.265pt}}
\multiput(475.00,382.92)(0.539,-0.497){47}{\rule{0.532pt}{0.120pt}}
\multiput(475.00,383.17)(25.896,-25.000){2}{\rule{0.266pt}{0.400pt}}
\multiput(502.00,357.92)(0.564,-0.496){43}{\rule{0.552pt}{0.120pt}}
\multiput(502.00,358.17)(24.854,-23.000){2}{\rule{0.276pt}{0.400pt}}
\multiput(528.00,334.92)(0.651,-0.496){37}{\rule{0.620pt}{0.119pt}}
\multiput(528.00,335.17)(24.713,-20.000){2}{\rule{0.310pt}{0.400pt}}
\multiput(554.00,314.92)(0.713,-0.495){35}{\rule{0.668pt}{0.119pt}}
\multiput(554.00,315.17)(25.613,-19.000){2}{\rule{0.334pt}{0.400pt}}
\multiput(581.00,295.92)(0.768,-0.495){31}{\rule{0.712pt}{0.119pt}}
\multiput(581.00,296.17)(24.523,-17.000){2}{\rule{0.356pt}{0.400pt}}
\multiput(607.00,278.92)(0.849,-0.494){29}{\rule{0.775pt}{0.119pt}}
\multiput(607.00,279.17)(25.391,-16.000){2}{\rule{0.388pt}{0.400pt}}
\multiput(634.00,262.92)(0.873,-0.494){27}{\rule{0.793pt}{0.119pt}}
\multiput(634.00,263.17)(24.353,-15.000){2}{\rule{0.397pt}{0.400pt}}
\multiput(660.00,247.92)(1.012,-0.493){23}{\rule{0.900pt}{0.119pt}}
\multiput(660.00,248.17)(24.132,-13.000){2}{\rule{0.450pt}{0.400pt}}
\multiput(686.00,234.92)(1.052,-0.493){23}{\rule{0.931pt}{0.119pt}}
\multiput(686.00,235.17)(25.068,-13.000){2}{\rule{0.465pt}{0.400pt}}
\multiput(713.00,221.92)(1.099,-0.492){21}{\rule{0.967pt}{0.119pt}}
\multiput(713.00,222.17)(23.994,-12.000){2}{\rule{0.483pt}{0.400pt}}
\multiput(739.00,209.92)(1.251,-0.492){19}{\rule{1.082pt}{0.118pt}}
\multiput(739.00,210.17)(24.755,-11.000){2}{\rule{0.541pt}{0.400pt}}
\multiput(766.00,198.92)(1.203,-0.492){19}{\rule{1.045pt}{0.118pt}}
\multiput(766.00,199.17)(23.830,-11.000){2}{\rule{0.523pt}{0.400pt}}
\put(299,647){\circle{24}}
\put(475,384){\circle{24}}
\put(638,281){\circle{24}}
\put(715,200){\circle{24}}
\put(299.0,624.0){\rule[-0.200pt]{0.400pt}{11.081pt}}
\put(289.0,624.0){\rule[-0.200pt]{4.818pt}{0.400pt}}
\put(289.0,670.0){\rule[-0.200pt]{4.818pt}{0.400pt}}
\put(475.0,376.0){\rule[-0.200pt]{0.400pt}{3.854pt}}
\put(465.0,376.0){\rule[-0.200pt]{4.818pt}{0.400pt}}
\put(465.0,392.0){\rule[-0.200pt]{4.818pt}{0.400pt}}
\put(638.0,264.0){\rule[-0.200pt]{0.400pt}{8.191pt}}
\put(628.0,264.0){\rule[-0.200pt]{4.818pt}{0.400pt}}
\put(628.0,298.0){\rule[-0.200pt]{4.818pt}{0.400pt}}
\put(715.0,157.0){\rule[-0.200pt]{0.400pt}{20.476pt}}
\put(705.0,157.0){\rule[-0.200pt]{4.818pt}{0.400pt}}
\put(705.0,242.0){\rule[-0.200pt]{4.818pt}{0.400pt}}
\end{picture}
\end{center}
\caption{Values of the QCD coupling constant~$\alpha_P$ determined from
the plaquette in simulations with differing lattice
spacings corresponding to $\beta\!=\!5.7$, 6, 6.2 and~6.4, all 
with $n_f\!=\!0$. The
coupling constant is plotted versus the average momentum $q_{1,1}$ carried 
by gluons in the plaquette at the various lattice spacings, 
with $q_{1,1}\!=\!3.4/a$. The line shows the coupling constant evolution 
predicted by third-order perturbation theory.}
\label{evol-fig}
\end{figure}

If one groups the various couplings in this table according to the
splitting used to tune the simulation and the number of light-quark
flavors~$n_f$, one finds that the values within a single group are
completely consistent. In particular, results obtained
using different loops are in excellent agreement, which shows that our
estimates of the errors caused by truncating perturbation theory are
reasonable. Also, the coupling constants obtained from the plaquette using 
$\beta$'s ranging from~5.7 to~6.4, corresponding to scales $q_{1,1}$ ranging
from 4.8\,GeV to 14.2\,GeV, agree well. This demonstrates that the
evolution of our coupling constant~$\alpha_\Psub$ is well described by
the perturbative beta function; no lattice artifacts are apparent. 
This is also
illustrated by Figure~\ref{evol-fig}, where we plot the
coupling constant~$\alpha_P(q_{1,1})$, obtained from the plaquette,
versus the effective momentum scale $q_{1,1}\!=\!3.4/a$ at which the
coupling is measured on each lattice. The simulation results for the
running of $\alpha_\Psub$ agree well
with the prediction of third-order perturbation theory\,\cite{running-alpha}.  

\subsection{Extrapolation to $n_f\!=\!3$}
The coupling constants in Table~\ref{alphaP-evol-table} from simulations
with different $n_f$'s are significantly different, as are the couplings
from simulations tuned using different meson mass splittings. Our final
step is to extrapolate to $n_f\!=\!3$, which is the correct number of
light-quark flavors for $\Upsilon$ and $\psi$ physics. The extrapolated
results, which are shown in Table~\ref{alphaP-extrap-table}, should
all agree, and do. To make the extrapolation, we paired  $n_f\!=\!0$ and
$n_f\!=\!2$ simulations as indicated in the table. 
For each separate combination of Wilson loop and meson mass
splitting, we
extrapolated $1/\alpha_\Psub$ using the corresponding
$\alpha_\Psub$'s from the two simulations.

\begin{table}
\begin{center}
\begin{tabular}{cclll}
\hline \\
$\beta$  & loop &
\multicolumn{3}{c}{$\alphathree_\Psub(8.2\,{\rm GeV})$} 
\\ \cline{3-5}
&& \multicolumn{1}{c}{$\PS$}   & 
\multicolumn{1}{c}{$\SS$}  & 
\multicolumn{1}{c}{$\PSc$} \\[1.5ex]
\hline 
6.0,5.6 & $-\ln W_{1,1}$ &  .1946\,(41)(0) & .1960\,(61)(0)\\
& $-\ln W_{1,2}$ &  .1913\,(42)(52) & .1927\,(58)(54) \\
& $-\ln W_{1,3}$  & .1897\,(42)(69)& .1912\,(58)(71) \\
& $-\ln W_{2,2}$ &  .1889\,(40)(120) & .1903\,(57)(125)\\
\\
5.7,5.415 & $-\ln W_{1,1}$ & .1884\,(57)(0) & & .1841\,(146)(0) \\
\hline
\end{tabular}
\end{center}
\caption{Values of $\alphathree_\Psub(8.2\,{\rm GeV})$ from different
operators and different tunings of the QCD simulation. The
two uncertainties listed are due to uncertainties in the inverse
lattice spacing, and to truncation errors in the extraction of
$\alpha_\Psub$ using perturbation theory. }
\label{alphaP-extrap-table}
\end{table}

We chose to extrapolate $1/\alpha_\Psub$ rather than $\alpha_\Psub$
because numerical experiments using third-order perturbation theory
suggest that $1/\alpha_\Psub$ is significantly more linear in~$n_f$. To
see how the couplings from our simulations might depend on $n_f$, note
that the $\Upsilon$~splittings that we use to determine the lattice spacing
probe QCD at momentum scales $q_\Upsilon$ of the order 0.5--1\,GeV.
Thus when we choose a lattice spacing that gives these splittings their
correct physical values, we are in effect tuning the QCD coupling
constant in our simulation to have its correct value at the 
scale~$q_\Upsilon$.  (If $n_f\!\neq\!3$, the simulation's coupling will 
have the correct value {\em only} at~$\qups$.) 
This means that the couplings in our
$n_f\!=\!0$ and 2 simulations agree with the correct $n_f\!=\!3$ coupling
at $q_\Upsilon$:
\be \label{alphas-equal}
\alphazero_\Psub(\qups) = \alphatwo_\Psub(\qups) =
\alphathree_\Psub(\qups).
\ee

This equation specifies the dependence of the couplings obtained in our 
simulations on~$n_f$, but we are unable to use it directly since perturbation
theory is not particularly reliable at~$\qups$. Nevertheless, we can use
this relation to test different schemes to extrapolate~$n_f$ as follows.  Taking
$\qups\!=\!1$\,GeV, we set all the couplings at that scale
equal to some large value, say .65. We then evolve all three to~8.2\,GeV
using the three-loop  beta function. Finally, we compare the
8.2\,GeV~coupling extrapolated from $n_f\!=\!0$ and 2 with the $n_f\!=\!3$
coupling obtained by evolving from~$\qups$. Extrapolating $\alpha_\Psub$
gives results that are ``correct'' to within 1.4\%, while
extrapolating $1/\alpha_\Psub$ is correct to within 0.3\%.
This exercise indicates that we should extrapolate the inverse coupling
and that the extrapolation errors are probably less than 1\%. Such errors 
are negligible relative to the other systematic and statistical
errors. Nevertheless, it would be desirable to repeat our analysis using
simulations with $n_f\!=\!3$ or even $n_f\!=\!4$.

\eq{alphas-equal} played a key role in the earliest determinations of the
running coupling constant using lattice QCD\,\cite{early-alphas}. These
studies used only $n_f\!=\!0$ simulations.  As can be seen from our
results, the coupling at $n_f\!=\!0$ is 25\% smaller than the correct
$n_f\!=\!3$ coupling. This correction was estimated in these earlier papers by
perturbatively evolving the $n_f\!=\!0$ coupling down to $\qups$, changing $n_f$ to
three, and then evolving back up to the original large scale, which is 8.2\,GeV in
the present analysis. This procedure suggests a correction of 15--20\%, which our
simulations show to be an underestimate but within the error range
quoted in the earlier papers.  We emphasize that there is no inconsistency
between these earlier analyses and ours. Our simulations with $n_f\!=\!0$
give results that are identical with the earlier work. What is different
here is that we have actual simulation results at $n_f\!\ne\!0$ and so get
to $n_f\!=\!3$ using extrapolation, rather than a perturbative
analysis that is well-motivated but only partly justified. 
That the sizable correction due to light-quark vacuum
polarization was so accurately predicted using perturbation theory
strengthens our confidence that our nonperturbative treatment of 
vacuum polarization is correct.  Note that if we
use the perturbative analysis to correct just our $n_f\!=\!2$ couplings,
ignoring our $n_f\!=\!0$ couplings, we obtain results that are in
excellent agreement with the extrapolated coupling\,\cite{japan-group}.

Our final results for $\alpha_\Psub$ in Table~\ref{alphaP-extrap-table}
agree well with each other and with our earlier results\,\cite{davies95}. 
In particular, the $5\,\sigma$ discrepancy between results using different
$\Upsilon$~splittings at $n_f\!=\!0$ disappears completely at $n_f\!=\!3$.
This is highly nontrivial; we are in effect counting the
number of light-quark flavors that affect real upsilons. It provides
confirmation that the quark vacuum polarization is correctly included in our
simulations and extrapolation.

\subsection{Conversion to $\alpha_\msbar$}
To compare with nonlattice determinations of the coupling constant, we
have converted our results to the $\MSbar$~definition of the coupling,
using \eq{msb-v} with $X_\msbar\!=\!.95\pm.95$. Our results are listed in
Table~\ref{alpha3msb-table}, and together with our~$\alpha_\Psub$'s in
Table~\ref{alphaP-extrap-table}, are the main result of this paper.
The $\MSbar$~results are somewhat larger than in our earlier paper because
we now use the $n_f\!=\!0$~value for $X_\msbar$, rather than setting it to
zero as before. Our estimate in the earlier paper for the size of
this term was correct and was included as an error. Consequently, our
old results are consistent with our new results within errors.

\begin{table}
\begin{center}
\begin{tabular}{cclll}
\hline \\
$\beta$  & loop &
\multicolumn{3}{c}{$\alphathree_\msbar(3.56\,{\rm GeV})$} 
\\ \cline{3-5}
&& \multicolumn{1}{c}{$\PS$}   & 
\multicolumn{1}{c}{$\SS$}  & 
\multicolumn{1}{c}{$\PSc$} \\[1.5ex]
\hline 
6.0,5.6 & $-\ln W_{1,1}$ &  .2258\,(56)(74) & .2277\,(83)(75) \\
& $-\ln W_{1,2}$ &  .2213\,(56)(99) & .2232\,(79)(102) \\
& $-\ln W_{1,3}$  & .2192\,(57)(116) & .2212\,(79)(119) \\
& $-\ln W_{2,2}$ &  .2181\,(54)(176) & .2200\,(77)(183) \\
\\
5.7,5.415 & $-\ln W_{1,1}$ &.2174\,(76)(67) & & .2117\,(197)(62) \\
\hline
\end{tabular}
\end{center}
\caption{Values of $\alphathree_\msbar(3.56\,{\rm GeV})$ from different
operators and different tunings of the QCD simulation. The
two uncertainties listed are due to uncertainties in the inverse
lattice spacing, and to truncation errors in the extraction of
$\alpha_\Psub$ and conversion to $\alpha_\msbar$ using perturbation theory. }
\label{alpha3msb-table}
\end{table}

To further facilitate comparisons with other analyses, we have numerically
integrated the third-order perturbative evolution equation for
$\alpha_\msbar$ and applied appropriate matching conditions at quark
thresholds\,\cite{matching-paper} to evolve it to the mass of the $Z^0$.
The results for our ten determinations are shown in
Table~\ref{alphamsbar-mz-table}. For matching we assumed $\MSbar$ masses
of 1.3(3)\,GeV and 4.1(1)\,GeV for the
$c$ and $b$ quarks respectively\,\cite{davies94,matching-paper}. The
uncertainties in these masses can shift the final coupling constant by
less than half a percent; we ignore them.

\begin{table}
\begin{center}
\begin{tabular}{cclll}
\hline \\
$\beta$  & loop &
\multicolumn{3}{c}{$\alphafive_\msbar(M_Z)$} 
\\ \cline{3-5}
&& \multicolumn{1}{c}{$\PS$}   & 
\multicolumn{1}{c}{$\SS$}  & 
\multicolumn{1}{c}{$\PSc$} \\[1.5ex]
\hline 
6.0,5.6 & $-\ln W_{1,1}$ & .1174\,(15)(19) & .1180\,(22)(20) \\
& $-\ln W_{1,2}$ & .1163\,(15)(26) & .1168\,(21)(27)\\
& $-\ln W_{1,3}$  & .1157\,(15)(31) & .1162\,(21)(31) \\
& $-\ln W_{2,2}$ &  .1154\,(14)(46) &  .1159\,(20)(48) \\
\\
5.7,5.415 &$-\ln W_{1,1}$ & .1152\,(20)(18) & & .1136\,(52)(16) \\
\hline
\end{tabular}
\end{center}
\caption{Values of $\alphafive_\msbar(M_Z)$ from several
operators and various tunings of the QCD simulation. The
two uncertainties listed are due to uncertainties in the inverse
lattice spacing, and to truncation errors in perturbative
expansions. }
\label{alphamsbar-mz-table}
\end{table}

\section{Discussion and Conclusions}

In this paper we have demonstrated that lattice simulations provide among
the simplest, most accurate, and most reliable determinations of the
strong coupling constant. Our ten different results, tabulated in
Tables~\ref{alphaP-extrap-table}--\ref{alphamsbar-mz-table}, are in
excellent agreement with each other. Indeed, all but one of them agree
with our best determination to within {\em its} uncertainty; that is,
to within the smallest error bars. Our best result implies
 \be
 \alphanf_\msbar(Q) = \cases{
 0.3706\,(288) & \mbox{for $Q=1.3~\GeV\approx M_c$ and $n_f=3$}\cr
 0.3701\,(288) & \mbox{for $Q=1.3~\GeV\approx M_c$ and $n_f=4$}\cr \cr
 0.2234\,(93)  & \mbox{for $Q=4.1~\GeV\approx M_b$ and $n_f=4$}\cr
 0.2233\,(93)  & \mbox{for $Q=4.1~\GeV\approx M_b$ and $n_f=5$}\cr \cr
 0.1174\,(24)  & \mbox{for $Q=91.2~\GeV = M_Z$ and $n_f=5$} \, ,\cr}
 \ee
with errors due to lattice-spacing and perturbation-theory 
uncertainties combined in quadrature.
These results are about $1\,\sigma$ higher than our previous
results\,\cite{davies95}. The shift is entirely due to the new
third-order term in the perturbative 
formula, \eq{msb-v}, relating the lattice coupling
$\alpha_\Psub$ to $\alpha_\msbar$. Our Monte Carlo simulation results
are essentially identical to those in our earlier paper. The shift
relative to our earlier result is only $1\,\sigma$ because we previously
estimated the size of this third-order term accurately.

The bulk of our effort in this analysis was devoted to understanding
and estimating the systematic errors. We varied every parameter in the
simulation. We used four different short-distance quantities to
extract the coupling, and three different (infrared) meson splittings,
in two different meson families, to tune
the bare coupling or lattice spacing. We demonstrated that the gross
features of $\Upsilon$ and $\psi$ physics are accurately described by
our simulations. We explored the role of light-quark vacuum
polarization for a range of light-quark masses. Our simulations were
sufficiently accurate to show that $n_f\!=\!0$ is the wrong number of
light-quark flavors for $\Upsilon$'s.  Only when we extrapolated to
$n_f\!=\!3$, the correct value, did our simulation results agree with
experiment. To see how robust
our results are, we redid the analysis but with various
ingredients missing. The corresponding shifts in
$\alphafive_\msbar(M_Z)$ are listed in Table~\ref{mutilation-table};
omitting the $n_f$ extrapolation led to the only appreciable difference.

\begin{table}
\begin{center}
\begin{tabular}{lc}
\hline 
& $\Delta \alphafive_\msbar(M_Z)$ \\ \hline 
omit $\order(a^2)$ gluonic corrections & $-0.6$\% \\
omit tadpole improvement of NRQCD & $-0.5$\% \\
omit $\order(v^2,a,a^2)$ corrections in NRQCD & $+0.9$\% \\
omit extrapolation (use $n_f\!=\!2$) & $-4.7$\% \\
\hline
\end{tabular}
\end{center}
\caption{Changes in the coupling constant at $M_Z$ when different
parts of our simulation or analysis are omitted.}
\label{mutilation-table}
\end{table}

The various parts of our analysis agree well with the results of other
groups. The $\alpha_\Psub$'s that we extract from Wilson loop
operators agree to within statistical and truncation errors with those
obtained by very different techniques\,\cite{weisz96}. This is the
easy part of the analysis. The remainder, involving the determination
of lattice spacings, has now also been duplicated. A
recent analysis of simulation results from the Fermilab and SCRI
groups, both of which employ a totally different formalism for
$b$-quark dynamics, gives $\alphafive_\msbar(M_Z) = .116\,(3)$, in
complete agreement with our results\,\cite{junko}.

Lattice coupling constant determinations such as ours enjoy a 
fundamental advantage over traditional methods 
based on perturbative high-energy  processes, allowing
significantly greater accuracy. 
The systematic uncertainties in the perturbative parts of the analyses
are similar in both approaches, but the nonperturbative elements
differ substantially.  When we tune our simulation to reproduce
the $\Upsilon$~spectrum, we are in effect directly tuning the QCD scale
parameter~$\Lambda_\msbar$. Consequently, a 5\% simulation error in a
mass splitting results in a 5\% error in $\Lambda_\msbar$, which
implies only a 1\% error in $\alpha_\msbar(M_Z)$. In high-energy
determinations, however, one measures the coupling constant rather than the 
scale parameter, and usually only through small radiative corrections
to an electroweak process.  Measuring~$\Lambda_\msbar$ is
intrinsically much more accurate than measuring~$\alpha_\msbar$.

There are prospects for substantially improving the accuracy of our
result fairly soon.  We list sources of error in our value for
$\alphafive_\msbar(M_Z)$ in Table~\ref{error-table}.  The
dominant error is due to truncation in perturbative expansions, specifically
because the $n_f$~dependent parts of our third-order coefficients have not 
yet been calculated. 
The agreement we observe between couplings from different loop
operators, each with its own perturbative series, suggests that our
estimates of this systematic error are realistic or even pessimistic.
Nevertheless, our total error could be cut in half by computing this
$n_f$ dependence, particularly for \eq{msb-v}. 
This is a straightforward perturbative
calculation. For this paper, we halved our statistical errors for
our $n_f\!=\!0$ simulations; the same should be done for
$n_f\!\ne\!0$. 
Use of an improved gluon action would remove the need for the
$a^2$ correction in the $\PS$ analysis, while it already has negligible effect on
$\SS$. The additional cost would be small~\cite{coarse-lattice}. 
Using a relativistic formulation of $c$-quark dynamics, rather than
NRQCD, might allow accurate results from the charmonium spectrum.
A simulation with either $n_f=3$ or $4$ light quarks would
eliminate the extrapolation error and would require perhaps only twice
the computational effort needed for $n_f=2$. 
Finally, simulations with {\em larger}
light-quark masses~$m_\eff$ would allow us to pin down more accurately
the dependence on this parameter. 
\begin{table}
\begin{center}
\begin{tabular}{lr}
\hline
Source & Uncertainty \\
\hline\\[-.3cm]
Unknown $n_f$ dependence in third-order perturbation theory &  1.9\%  \\[.1cm]
Statistical error in determination of $a^{-1}$           &  .9\%  \\[.1cm]
Light-quark masses                                       &  .9\% \\[.1cm]
Extrapolation in $n_f$                                   &  .3\% \\[.1cm]
Finite $a$ and $\order(v^4)$ errors                      &  .2\% \\[.1cm]
Fourth-order evolution of $\alpha_\msbar$                &  .01\% \\[.05cm]
\hline
\end{tabular}
\end{center}
\caption{Sources of error in our best determination
of $\alphafive_\msbar(M_Z)$.}
\label{error-table}
\end{table}

Our lattice determinations of the strong coupling constant agree well
with most determinations based on  perturbative high-energy
processes. This fact provides striking evidence that the
nonperturbative QCD of hadronic confinement and the perturbative QCD
of high-energy jets are the same theory.

\section*{Acknowledgements}
We thank Urs Heller, Aida El-Khadra, Martin L\"uscher, Paul Mackenzie,
Chris Michael, and Peter Weisz 
for several useful conversations. We also thank Andrew Lidsey for his
contribution to our $\beta\!=\!5.7$~analysis. This
work was supported in the U.~K.\ by a grant from PPARC, and in the
U.~S.\ by grants from the National Science Foundation and the
Department of Energy.


\begin{thebibliography}{10}

\bibitem{susy} See, for example,
P.~N.~Burrows, {\em Review of $\alpha_s$ Measurements}, invited talk
at the Third International Symposium on Radiative Corrections, Cracow,
Poland, August 1996.

\bibitem{davies95} C.~T.~H.~Davies, K.~Hornbostel,
G.~P.~Lepage,  A.~Lidsey, J.~Shigemitsu and J.~Sloan,
Phys.\ Lett.\ {B345}, 42 (1995).
Our formula for $\ln W_{1,1}$ is slightly different
from that in our earlier paper because we have shifted the
$\alpha_P$~scale from $3.41/a$ to $3.40/a$.  The latter
is the correct scale at infinite volume.

\bibitem{san-diego} An independent analysis of this effect that uses
preliminary versions of our data is given in 
B.~Grinstein and I.~Z.~Rothstein, 
Phys.\ Lett.\ {\bf B385}, 265 (1996).

\bibitem{s-mass-papers}
See, for example,
R.~Gupta and T.~Bhattacharya, Los Alamos preprint  LA-UR-96-1840
(May 1996), hep-lat/9605039;
B.~J.~Gough, G.~M.~Hockney, A.~X.~El-Khadra, A.~S.~Kronfeld, 
P.~B.~Mackenzie, B.~P.~Mertens, T.~Onogi, J.~N.~Simone, Fermilab preprint 
FERMILAB-PUB-96-283-T (Oct 1996), hep-ph/9610223. 


\bibitem{nrqcd-papers} 
G.~P.~Lepage and B.~A.~Thacker,
Nucl.\ Phys.\ {\bf B} (Proc.\ Suppl.) {\bf 4} 199, (1988);
B.~A.~Thacker and G.~P.~Lepage, Phys.\ Rev.\ D {\bf 43}, 196 (1991);
G.~P.~Lepage,  L.~Magnea,  C.~Nakhleh, U.~Magnea and K.~Hornbostel,
Phys.\ Rev.\ D {\bf 46}, 4052 (1992).

\bibitem{davies-upsilon} C.~T.~H.~Davies, K.~Hornbostel, A.~Langnau,
G.~P.~Lepage,  A.~Lidsey, J.~Shigemitsu and J.~Sloan,
Phys.\ Rev.\ D {\bf 50}, 6963 (1994).


\bibitem{davies-psi}
C.~T.~H.~Davies, K.~Hornbostel, G.~P.~Lepage, A.~J.~Lidsey, 
J.~Shigemitsu and J.~Sloan, 
Phys.\ Rev.\ {\bf D52} 6519 (1995). 

\bibitem{davies94}
C.~T.~H.~Davies, K.~Hornbostel, A.~Langnau, G.~P.~Lepage, A.~J.~Lidsey, 
C.~Morningstar, J.~Shigemitsu and J.~Sloan, 
Phys.\ Rev.\ Lett.\ {\bf 73} 2654 (1994).

\bibitem{kilcup} These gauge-field configurations were provided by G.~Kilcup
and collaborators. There were 105~configurations on $16^3\times24$~lattices.

\bibitem{kogut} These gauge-field configurations were provided by 
J.~Kogut and collaborators. There were
149~configurations on $16^3\times32$~lattices. 

\bibitem{ukqcd} These gauge-field configurations were provided by the
UKQCD~collaboration. There were 200~configurations on
$12^3\times24$~lattices for $\beta\!=\!5.7$, and 216~configurations
on $24^3\times48$~lattices for $\beta\!=\!6.2$.


\bibitem{hemcgc} The gauge-field configurations were provided by the
HEMCGC collaboration, K.~Bitar {\it et al.}, Nucl.\ Phys.\ {\bf B {\rm
(Proc.\ Suppl.)} 26}, 259 (1992); Phys.\ Rev.\ D~{\bf 46}, 2169 (1992);
Phys.\ Rev.\ D~{\bf 48}, 370 (1993).
There were 399~configurations at $am\!=\!.01$ and 199~at
$am\!=\!.025$, all on $16^3\times32$~lattices. 
The staggered quark action was used for the light quarks, and the 
Hybrid Molecular Dynamics algorithm for the dynamical configurations.

\bibitem{milc} The gauge-field configurations were provided by the
MILC collaboration,
C.~Bernard {\it et al}, Nucl.\ Phys.\ B (Proc.\ Suppl.) {\bf30}, 369(1996);
Phys.\ Rev.\ {\bf D48}, 4419 (1993); Nucl.\ Phys.\ B (Proc.\ Suppl.) {\bf34},
366 (1994); Phys.\ Rev.\ {\bf D45}, 3854 (1992).
There were 299 configurations at $am\!=\!.0125$ on $16^3\times32$~lattices.
The staggered quark action was used for the light quarks, and the 
Hybrid Molecular Dynamics algorithm for the dynamical configurations.

\bibitem{cambridge} 
S.~M.~Catterall, F.~R.~Devlin, I.~T.~Drummond and R.~R.~Horgan,
Phys.\ Lett.\  {\bf 321B}, 246 (1994).

\bibitem{trottier}
H.~Trottier, Simon Fraser preprint SFU-HEP-131-96 (Nov 1996),
hep-lat/9611026.

\bibitem{yndurain} 
F.~J.~Yndurain, {\it The Theory of Quark and Gluon Interactions}, 
Springer (1993).

\bibitem{lm} 
G.~P.~Lepage and P.~B.~Mackenzie, Phys.\ Rev.\ D~{\bf 48}, 2250 (1993). 

\bibitem{blm} 
S.~J.~Brodsky, G.~P.~Lepage and P.~B.~Mackenzie, Phys.\ Rev.\
D~{\bf 28}, 228 (1983). 

\bibitem{wloop-pth-ref} The second-order coefficients were obtained by
extrapolating to infinite volume those found in U.~Heller and F.~Karsch, Nucl.
Phys. {\bf B251}, 254 (1985); Nucl.\ Phys.\ {\bf B258}, 29 (1985);
H.~Hamber and C.~Wu, Phys.\ Lett.\ {\bf 127B}, 119 (1983).
Third-order coefficients are from Monte Carlo simulations, W.~Dimm,
G.~Hockney, G.~P.~Lepage and P.~B.~Mackenzie, unpublished. These have a
similar uncertainty. See also P.~Weisz, Phys.\ Lett.\ {\bf 100B}, 331
(1981); H.~S.~Sharatchandra, H.~J.~Thun and P.~Weisz, Nucl.\ Phys.\ {\bf
B192}, 205 (1981); K.~M.~Bitar {\it et al}, Phys.\ Rev.\ D~{\bf
48}, 370 (1993).


\bibitem{luscher95} This coefficient is obtained by combining the
definition of $\alphaP$ (see \cite{davies95}) with results in
B.~Alles, M.~Campostrini, A.~Feo,
H.~Panagopoulos, Phys.\ Lett.\ {\bf B324}, 433 (1994); 
M.~Luscher and P.~Weisz,
Phys.\ Lett.\ {\bf B349}, 165 (1995); 
Nucl.\ Phys.\ {\bf B452}, 234 (1995). 

\bibitem{markus} 
M.~Peter, Phys. Rev.\ Lett.\ {\bf78} 602 (1997).

\bibitem{appelquist} T.~Appelquist, M.~Dine,
I.~J.~Muzinich, Phys.\ Rev.\ {\bf D17}, 2074 (1978);
Phys.\ Lett.\ {\bf69B}, 231 (1977).

\bibitem{wloops-mc} The $n_f\!=\!0$ Monte Carlo results are from the
Fermilab study discussed in Ref.~\cite{lm}; those for $n_f\!=\!2$ are
from~\cite{hemcgc} and~\cite{milc}.

\bibitem{running-alpha} See \cite{lm} and also 
S.~P.~Booth, D.~S.~Henty, A.~Hulsebos, A.~C.~Irving, C.~Michael and
P.~W.~Stephenson,
Phys.\ Lett.\ {\bf 294B}, 385 (1992);
M.~L\"uscher, R.~Sommer, P.~Weisz and U.~Wolff,
Nucl.\ Phys.\ {\bf B413}, 481 (1994); 
K.~Schilling and G.~S.~Bali, 
Nucl.\ Phys.\ {\bf B} (Proc.\ Suppl.) {\bf 34}, 147 (1994);
and references therein.

\bibitem{early-alphas} A.~X.~El-Khadra, G.~Hockney,
A.~S.~Kronfeld, P.~B.~Mackenzie,  Phys.\ Rev.\ Lett.\ {\bf 69}, 729 (1992); 
A.~X.~El-Khadra, Nucl.\ Phys.\ B (Proc.\ Suppl.) {\bf 34}, 141 (1994);
G.~P.~Lepage, Nucl.\ Phys.\ {\bf B} (Proc.\ Suppl.) {\bf 26}, 45 (1992);
T.~Onogi, S.~Aoki, M.~Fukugita, S.~Hashimoto, N.~Ishizuka, 
H.~Mino, M.~Okawa, A.~Ukawa,
Nucl.\ Phys.\ {\bf B} (Proc.\ Suppl.) {\bf 34}, 492 (1994).

\bibitem{japan-group} 
Another determination based upon an $n_f\!=\!2$
simulation and a perturbative correction for $n_f$ is given in
S.~Aoki, M.~Fukugita, S.~Hashimoto, N.~Ishizuka, H.~Mino, 
M.~Okawa, T.~Onogi, and A.~Ukawa, Phys.\ Rev.\ Lett.\ {\bf 74}, 22 (1995). 
Their results agree with ours, within larger errors.


\bibitem{matching-paper} 
For a useful discussion of this procedure, see
G.~Rodrigo and A.~Santamaria, Phys.\ Lett.\ {\bf 313B},
441 (1993).  

\bibitem{quark-masses} The $b$-quark mass is from~\cite{davies94}.

\bibitem{weisz96} P.~Weisz, Nucl.\ Phys.\ B (Proc.\ Suppl.) {\bf 47}, 71 (1996).

\bibitem{junko} 
J.~Shigemitsu, 
Nucl.\ Phys.\ B (Proc.\ Suppl.) {\bf 53}, 16 (1997).

\bibitem{coarse-lattice}
M.~Alford, W.~Dimm, G.~P.~Lepage, G.~Hockney, P.~B.~Mackenzie,
Phys.\ Lett.\ {\bf B361}, 87 (1995). 

\end{thebibliography}
\end{document}